# A Language for Large-Scale Collaboration in Economics: A Streamlined Computational Representation of Financial Models

Jorge M. Faleiro Jr. [#1]

[#] *Centre of Computational Finance and Economic Agents, University of Essex*
*Wivenhoe Park, Colchester, CO4 3SQ, UK*
[1] jfalei@essex.ac.uk    j@falei.ro

*Abstract*[1]— **This paper introduces *Sigma*, a domain-specific computational representation for collaboration in large-scale for the field of economics.**

**A computational representation is not a programming language or a software platform. A computational representation is a domain-specific representation system based on three specific elements: facets, contributions, and constraints of data. Facets are definable aspects that make up a subject or an object. Contributions are shareable and formal evidence, carrying specific properties, and produced as a result of a crowd-based scientific investigation. Constraints of data are restrictions defining domain-specific rules of association between entities and relationships.**

**A computational representation serves as a layer of abstraction that is required in order to define domain-specific concepts in computers, in a way these concepts can be shared in a crowd for the purposes of a controlled scientific investigation in large-scale by crowds.**

**Facets, contributions, and constraints of data are defined for any domain of knowledge by the application of a generic set of inputs, procedural steps, and products called a representational process.**

**The application of this generic process to our domain of knowledge, the field of economics, produces Sigma. Sigma is described in this paper in terms of its three elements: facets (streaming, reactives, distribution, and simulation), contributions (financial models, processors, and endpoints), and constraints of data (configuration, execution, and simulation meta-model).**

**Each element of the generic representational process and the Sigma computational representation is described and formalized in details.**

I.  Background

Languages are more than a vehicle for communication. They are often one's windows to reality. A language shapes how a person thinks, what can be achieved, and how can be achieved. Some languages often come from and facilitate the representation of concepts in a specific domain of knowledge, and if used outside of that specific domain, could make the representation of those same concepts more obscure. A person might, for example, use the German language for philosophy, or French for poetry. Using them the other way around might make a person write more, or being forcibly more verbose, or even lose clarity. In extreme cases using the wrong language for a domain of knowledge can impede the expression of the exact ideas one might intend.

Language defines reality [1] [2]. Human observations that lead to the scientific inquiry, and drive our process of discovery are shaped by our method of questioning, and limited by the language we possess [3][2].

This paper explores literature and evidence showing that this is not only the case with natural languages but also with a layer of abstraction that is required to define domain-specific concepts in computers, in a way that these concepts can be shared in a crowd. We are calling this conceptual abstraction a *computational representation*.

By definition, a computational representation is a representation system based on three specific elements: facets, contributions, and constraints of data. A computational representation serves as a layer of abstraction that is required in order to define domain-specific concepts in computers, in a way these concepts can be shared in a crowd for the purposes of a controlled investigation in large-scale by crowds [4] [5] [6].

According to the theory of enablers, a computational representation is a non-cognitive enabler of crowd-based scientific investigation [6]. Non-cognitive enablers relate to features that can be directly and purely mapped to a computational description. Cognitive enablers, on the other hand, relate to non-computational features associated with the subjective mechanisms of human understanding of what to consider knowledge and the underlying fabrics of large-scale collaboration. Cognitive enablers are not domain specific, and as a consequence should be the same regardless of the domain of knowledge under consideration [6].

---

[1] Large portions of this paper are reproduced as part of [6]

[2] "We have to remember that what we observe is not nature itself but nature exposed to our method of questioning. Our scientific work (…) consists in asking questions about nature in the language that we possess and trying to get an answer from experiments by the means that are at our disposal" – Werner Heisenberg [3]

Similarly to natural languages, a computational representation grows from the needs of a specialized domain, and therefore is better suited for use cases relevant to that specific domain. In some domains of knowledge, like architectural sciences, one would be more concerned about spaces, shapes, volumes or colors, and their relationships with a three-dimensional environment and the effect of the interaction of those concepts with humans. In legal sciences, one would be more concerned about possible associations between real-world entities, and rules defining their behavior and constraints for interaction. In some other domains, like bioinformatics, the ability to represent interconnected shapes and strings could be more relevant. In biophysics, it is essential to keep track of genotypical and phenotypical traits, and their relationships with encoded protein sequences with a vast number of possible combinations. In economics, our subject of concern, a researcher would be more interested in the way changes in quantitative measurements, over a time series, would affect the valuation.

A computational representation must mimic the inherently free flow of thoughts of the human mind and the speed of modern vehicles of collaboration, and therefore, by similarity, a computational representation must be fluid. In contradiction, computational artifacts, like programming languages and databases, are born out of strictly technical aspects of a problem and bred outside of concerns relevant to specific domains of knowledge. Only after definition, they are forcedly introduced for use and therefore not able to follow the free-flow of the evolution of ideas. Computational artifacts remain frozen to domain-specific requirements of that specific point in time when the introduction occurred. When requirements on that domain evolve to follow the increasing complexity of the problems at hand, those artifacts would no longer fit, or in a best case require an additional verbosity, sacrificing the proper semantics of communication.

In opposition to computational artifacts, a computational representation must be dynamic, able to adapt and evolve to solve new classes of problems and organize increasingly complex and powerful computing environments. These new classes of problems are different from the problems we had to deal with just a few years back [5]. They require the collaboration of multi-disciplinary specialists exchanging different types of artifacts that must be adequately described and tracked [5] [7]. An investigator must have adequate tools and methods to approach new problems correctly. On this sense, an adequate computational representation allows for the proper description and control of those tools and methods, allowing them to change in the face of new demands and be able to address new problems [8].

Unfortunately, the status quo in exploratory research in general, and in economics in particular, defines a different reality. The lack of adequate representation and an abundance of computational power allow, and unintentionally require, a potentially obfuscated representation of ways to transform and store data, yielding massive amounts of convoluted and dissociated information. This paradoxical condition entangling modern investigative procedures define a vicious circle. Uncontrolled methods require more computing power, which enables to transform more data, which as a consequence bring an incentive for uncontrolled models and increasing amounts of untraceable data. These, in turn, require more opaque techniques and computing resources to trace and decipher that data, the "informatics crisis" [5] [9]. This paper proposes a computational representation for the field of economics to allow breaking this never-ending feedback cycle. We are naming this computational representation *Sigma[3]*.

The way in which an investigator describes to an increasingly complex machine a method to resolve a problem plays a fundamental role in communication and collaboration, and as a consequence in the traceability of the process of investigation and discovery. The amount of data generated in modern investigative procedures as input and output cannot be represented to humans the same way as they are to computers [10]. To make research truly useful, we need human-friendly ways to visualize, track, store and understand the evidence. In addition to representing evidence, communicating methods and procedures must be regarded as of greater importance than explanatory texts and figures as experimental outputs [11] [12]. Representation of the methods by which we represent procedures of investigation cannot be addressed differently than other items that require human visualization and interpretation.

Given the intrinsic association of a computational representation to a domain of knowledge, it would be natural to expect that a computational representation could be derived from a domain of knowledge, given a set of well-defined inputs and general procedures. We are calling this organization of inputs and general procedures to produce a computational representation a *representational process*.

II. A GENERIC METHOD TO PRODUCE COMPUTATIONAL REPRESENTATIONS: THE REPRESENTATIONAL PROCESS

The outline of a representational process to define a computational representation for any domain of knowledge is described in Figure 1.

The outline defined in Figure 1 shows a set of two inputs and four distinct steps that are necessary to generate a computational representation composed of facets, contributions, and constraints of data. Arrows define the flow of data, and not control. As a consequence, arrows define dependencies for the execution of a given step.

The two inputs for a representational process, represented on the top of the diagram, are the entry points for the representational process. The inputs are a set of domain-specific cases of use and a computational taxonomy.

Domain-specific cases of use are a collection of exercises reflecting specific characteristics of concern in that domain of knowledge. The selection of cases of use should represent an overreaching and diverse sample of the main activities relevant to that domain of knowledge. Each case of use defines the domain-specific knowledge necessary for that specific scenario to be understood and executed.

---

[3] The name comes from the usual reference to summation in mathematics, from which we borrow a connotation of aggregation, or collaboration. Sigma is a computational representation to define exchangeable financial models, for the specific purpose of communication for participants in large scale for economics.

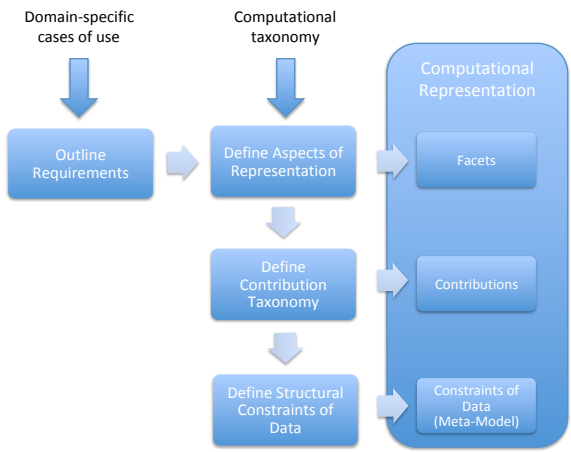

**Figure 1. Representational Process**

An outline of a generic process to define a computational representation for any domain of knowledge on four steps and two inputs: a list of exercises, or domain-specific cases of use, and a computational taxonomy.

A computational taxonomy is an inventory of computer technologies available and relevant for the implementation of cases of use at that moment in time. Examples of items in a computational taxonomy are technologies to store, retrieve, analyze, and visualize data and computational methods. A computational taxonomy is fluid, in a sense that the exact definition of what is relevant is affected by qualities of the individual using this process, such as experience, and personal biases. A discussion on the non-deterministic nature of the process, concerning a computational taxonomy, is given in Section III.

The four distinct steps of the representational process defined in Figure 1 are shown in individual solid boxes: outline requirements, define aspects of representation, define contribution taxonomy, and define structural constraints of data. Incoming arrows in each box define dependencies, and outgoing arrows define products, or results, of the execution of that specific step.

The first step outlines requirements that are relevant for the definition of a computational representation for a domain of knowledge. The outline of requirements is produced from a list of domain-specific cases of use, defined based on relevancy. Relevancy is given by, as we have mentioned before, the assumption that the set of cases of use is representative enough for most of the scenarios of investigation in that domain of knowledge. If the assumption is valid, we can infer as a consequence that any investigation exercise on that domain should depend, at least in a substantial part, with a combination of one or more of those requirements. For reasons of completeness, a proper computational representation for that domain of knowledge must address all these requirements. As a result, by definition, what we call a *proper computational representation* for a domain of knowledge should intend to represent all cases of use in the scope defined by the original list of cases of use[4].

---

[4] An example of an execution of the outline requirements step is given in the upcoming Section IV when we define requirements for a computational representation for the field of economics.

The second step definesf aspects of representation based on the computational taxonomy and the domain-specific requirements produced as a result of the first step. The result of the second step is the set of facets of a computational representation. An example of an execution of the step to define aspects of representation is given in the upcoming Section V when we describe facets and the process of their definition for a computational representation in economics.

The third step defines a contribution taxonomy based on facets produced as a result of the second step. The results of the third step are contributions of a computational representation. An example of the step to define contributions is given in the upcoming Section VI when we describe contributions and the process of their definition for a computational representation in economics.

The fourth step defines structural constraints of data based on facets and contributions produced as the result of the second and third step. The results of the fourth step are constraints of data, or meta-model, of a computational representation. An example of the step to define contributions is given in the upcoming Section VII when we describe constraints of data and the process of their definition for a computational representation in economics.

The final result of a representational process, as shown in Figure 1 by a larger solid box on the right side, is a computational representation given by facets produced in step two, contributions produced in step three, and constraints of data produced in step four. Each of the elements is depicted in Figure 1 as smaller boxes inside the computational representation. Facets, contributions, and constraints of data are detailed over the upcoming sections.

*A. Facets*

A facet, in the context of this research, is defined as "one of the definable aspects that make up a subject or an object; denomination of things that are similar or related, but yet distinct things" [13].

A more intuitive definition of what exactly is a facet is done by example and would come from a domain in which concepts are more tangible and organoleptic than in economics. Intuitiveness, as it is always the case, is achieved by representing concepts that are keen to one or more of traditional human senses.

Taking the domain of architecture, or civil engineering sciences, for example. The representation of ideas is done through the placement of volumetric shapes considering restrictions like light, gravity, and the mutually exclusive placement of objects in space. One example of a typical representation of that domain is shown in Figure 2.

Three-dimensional shapes, textures, colors, and measurements can be combined to define concepts like pieces of furniture, rooms, ambiance, and then extended to derive in computers notions that can only be asserted at naked eye, anticipating the effect of the interaction of these concepts with individuals.

These primary, fundamental elements that can be combined to generate core concepts on the domain of knowledge are called *facets*.

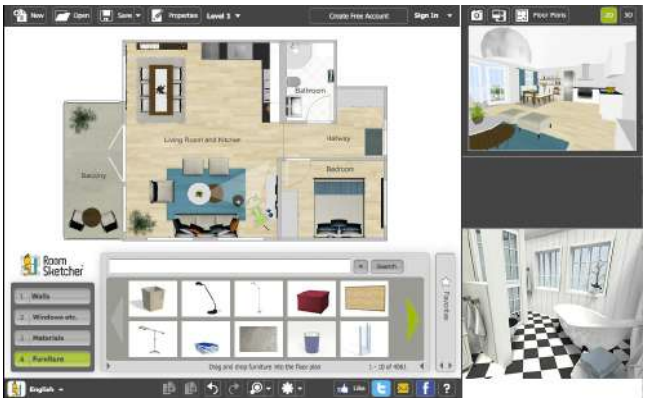

**Figure 2. Example of Facets in a Domain of Knowledge**

Aspects like volumetric shapes and specific coordinates in a 3D environment (facets) are used to describe a layout relevant to a specific domain of knowledge (architecture)

In this example shapes, texture, colors, and measurements are facets, representable in computers, which make up a subject or an object relevant to that specific domain of knowledge: architecture.

At this point, it is important to emphasize one of the core assumptions of this research: the exact definition of what constitutes a facet in a specific domain of knowledge is empirical. In some cases, e.g., our example related to architectural sciences, the proximity to visual and spatial concepts makes the establishment of what is indeed a facet - shapes, textures, color, and measurements - somewhat intuitive, and as a consequence more natural to derive[5].

### B. Contributions

In the scope of this research, we call contributions the set of shareable and formal evidence [6] of an objective investigation. As shareable evidence they can be exchanged, reused and traced through something called a record of provenance [7], therefore becoming a vehicle for effective collaboration.

To be qualified as contributions in a crowd-based investigation scenario, any evidence has to carry specific traits: evidential properties, intrinsicality, and characteristics of communication and interaction.

#### B.1. Evidential Properties

To be defined, shared, reused and traced contributions must carry particular mandatory traits we call evidential properties: classification, identification, a record of provenance, and ownership and security [7].

- **Classification**: Contributions must follow a classification system of shareable entities, specific to the domain of knowledge under consideration, and referred to as taxonomy of contributions. This classification system is an organization of shareable artifacts, organized based on relevant features[8].

- **Identification**: Contributions should be appropriately identified following common standards for shared identification in a way to allow reference, sharing, and ownership [14].

- **Provenance**: Contributions should carry a record of the chronology of ownership, custody or location of contributions, as well as the history of associations of contributions to entities or participants. We call this chronological description of custody and location a record of provenance.

- **Ownership and security**: Given the sensitive nature of contributions, contributions should ensure ownership and access only after proper authorization and authentication. For that reason, contributions must carry a record of ownership and authorization.

#### B.2. Intrinsicality

Contribution properties are defined as either intrinsic or extrinsic. Intrinsic properties of contributions [9] are not explicitly described in the representation and are enforced by an implementation of the computational platform. They can be assumed to be in place based on physical aspects of the contribution, regardless of specific indications on the representation. On the other hand, extrinsic properties are explicitly represented.

For example, ownership of each revision or improvement in a contribution occurs without an explicit description in a representation. As a consequence tracking the ownership of contributions occurs by the natural exchange of artifacts that are inherently traceable. In that way artifacts are traced when they are produced and utilized, making the record or provenance transparent [7]

#### B.3. Characteristics of Communication and Interaction

Contributions must carry characteristics to allow collaboration to take place. Collaboration is a direct result of how well contributions foster communication and interaction. A contribution must support three characteristics of communication and interaction to support large-scale collaboration: analytical description, granularity, and simplicity.

- **Analytical description**: Problems must be proposed in a way that allows for an analytical description, following a top-to-bottom structure. Splitting the description of problems into sub-tasks allows micro-expertise to be harnessed more directly and contributions to be naturally generated and associated with solutions.

- **Granularity**: A computational representation should encourage short, small contributions. Small contributions would make simpler and more straightforward for experts to review incoming collaboration and assess if they are relevant to their investigation.

---

[5] For the domain of knowledge of concern for this research, the selection of facets for a computational representation for the field of economics and their formalization is given in Section V.

[6] The available body of facts or information indicating whether a belief or proposition is true or valid [83]

[7] Chronology of the ownership, custody or location of historical entities [13]

[8] The classification of contributions for the domain of economics is depicted in Figure 15.

[9] Intrinsic elements of a representation are elements enforced by an implementation of the representation. An intrinsic element can be assumed to be in place regardless of any specific expressions on the representation itself.

- **Simplicity**: Representation of contributions should be simple and straightforward. A streamlined representation would make it easier to refer to foundational knowledge, as well as making it easier for participants to communicate and describe contributions.

These properties of analytical description, granularity and simplicity allow input and results from one experiment to be seamlessly utilized by other experiments, easing extensions on models and data to fit additional scenarios by short and specialized description.

The contribution taxonomy for the field of economics, listing the relevant properties for that specific domain of knowledge, is defined in Section VI.

### C. Constraints of Data

Most domains express real entities and relationships using structural constraints of data. Those constraints define rules of associations that establish what is feasible in that domain, in the real world.

These rules of associations define structural constraints of data in place for a specific domain of knowledge. Those structural constraints use an abstract layer of data to define restrictions on a separate layer of abstractions, based themselves on data, hence the term meta-data[10]. The set of structural constraints in a specific domain of knowledge is called meta-model.

Depending on the complexity of the domain of knowledge, and what should be represented, meta-data in a specific domain can follow a classification. A specific meta-model for the field of economics is described in Section VII.

### III. DISCUSSION ON ASSUMPTIONS AND CONSEQUENCES OF KNOWLEDGE REPRESENTATION THROUGH MODELS

The conceptual layout of a computational representation is, in essence, a proposal to represent knowledge in a given specialized field through abstractions commonly called *models*.

The representation of knowledge through models is not something new. There is a long history of academic work attempting similar tasks in a variety of domains [15] [16]. However, most works concentrate on a comparative analysis, evaluating properties of specific representations against others.

Alternatively, this research assumes a role-based definition of knowledge representation. In a role-based definition, a description of a knowledge system is defined in terms of five core roles a specific representation plays [17] [7].

- **Models are surrogates**: a surrogate is by definition a substitute for the target idea itself, and as a result, a measurement of how far or how close this surrogate is from calculations it intends to represent is secondary or irrelevant.

- **Models define human expressions**: models should define measurements and concepts understood by humans in a language that is adequate for human consumption, even if not directly natural.

- **Models are a medium for efficient computation**: models are a medium for pragmatic efficient computation, or in other words, models should be able to be replicated in computers given appropriate technology and sufficient resources.

- **Models establish ontological commitments:** models define ontological commitments for a representation by defining "a set of decisions about how and what to see in the world" [18] [19]. Models are approximations of reality, and as we define them, we make decisions of what to consider and what to ignore. These decisions are ontological commitments and are "not an incidental side effect but they are of essence in our representation" [17].

- **Models define a theory of intelligent reasoning**: Models define a "fragmentary theory of intelligent reasoning" represented in terms of concepts and inferences, sanctioned and recommended. Models represent "some insight indicating how people reason intelligently" about a problem or investigation [17].

The use of a role-based definition and these core roles bring important consequences when defining a computational representation for any domain of knowledge:

The first and most important consequence is that computational representations are non-discriminatory. In other words, computational representations should not be measured by how efficiently they represent a target idea, and therefore should not be compared to one another. Computational representations are abstract surrogates for a target idea, and as such, they are just a set of decisions of what to see in a subject, and therefore bound to limitations and biases of an observer.

Second, a computational representation and associated models are fluid and not final. To put differently, computational representations are not set in stone and are expected to change whenever noticeable changes in technology bring new methods and tools, or a new case of use becomes relevant for that specific domain of knowledge.

These assumptions and consequences are critical when assessing and understanding features and limitations of any computational representation defined from the representational process defined in Section II. These same assumptions and consequences should be expected in any representation, and more importantly, in the case of this research, in a computational representation for the field of economics.

### IV. DOMAIN-SPECIFIC REQUIREMENTS FOR ECONOMICS

A computational representation, as defined previously in Section I, is a representation system based on facets, contributions, and constraints of data and used to define concepts related to a specific domain of knowledge, in a way these concepts can be shared with a crowd to allow controlled investigation in large-scale.

---
[10] Data that provides information about other data [13]

A computational representation can be defined for any domain of knowledge by following the steps of the representational process defined in Section II. According to the representational process, a computational representation can be generated for any domain of knowledge given a set of domain-specific requirements and a computational taxonomy. The set of domain-specific requirements for the field of economics is defined over the following Section IV, and the computational taxonomy is presented as we describe each facet.

As explained in Section II, a computational representation is built based on a set of domain-specific requirements selected by careful examination of specific features of a number of domain-specific cases of use.

Each case of use defines the knowledge necessary for that specific scenario to be understood and executed. For the definition of domain-specific requirements that will be used for the definition of a computational representation for economics, each case of use is a separate empirical exercise:

- Assessing the performance of momentum cross-over strategies using Monte Carlo simulations and historical backtesting [8]
- Simulation of the performance of real-time strategies through backtesting [20]
- Profitability of different moving average cross-over strategies [21]
- Real-time valuation of an equities portfolio [22]
- Assessment of profitability of strategies holding long positions on fixed-length intervals [23].
- Agent-based simulation of a central limit order book [24] [25] [26] [27] [28] [29]

Some of these exercises are extensive and relate to novelty research subjects. Each one of those exercises expresses specific behaviors, later translated to an outline of features for proper representation of financial models, and as a consequence, requirements for a computational representation for the domain of economics. With that, the requirements for a computational representation for the field of economics as listed as follows:

- **Simplicity of communication**: a financial model is seldom defined and interpreted by one single group of users. The notation for its description should be simple enough to allow communication across a diverse community of users;
- **Predictability**: financial models are often defined with the intent of anticipating behavior or critical events;
- **Complexity of the domain of knowledge**: financial sciences deal with subjects that are inheritably complex and challenging to model;
- **Large volume of data**: virtually infinite history associated to a record of time: The record of the memory of financial models is associated with either datasets or streams of data that are virtually infinite.
- **Sliding window computations**: a sequence of fragments of data has to be evaluated so that adjacent members in the sequence, fitting a constant sliding time window, are relevant for the computation of a result[11].
- **Low latency**: responsiveness in near real-time. Given an event, or stimuli, some cases of use most respond as quickly as possible to avoid penalizing accuracy of measurements and profitability of the model itself;
- **Event-driven**: actions respond to events, originating from external and unpredictable sources;
- **Time-based**: tightly coupled with notions of value variations (e.g., prices, ratios) over discrete time series;
- **Graph-oriented**: financial models strongly rely on real-world entities and their ad-hoc relationships. Entities are associated with nodes and relationships to edges in graph-oriented representations. The sequence of transformations and steps to operate on real-world entities, either sequentially or not, is also graph oriented

We assume that the set of use cases is representative enough for most of the scenarios of investigation in economics. If the assumption is valid, we can infer as a consequence that any financial model should depend, at least in a substantial part, with a combination of one or more of those requirements. For reasons of completeness, a proper computational representation for the field of economics must address all these requirements. In this sense, by definition, a proper computational representational should intend to represent all cases of use in the scope of economics.

V. FACETS

The example provided previously in page 3, when we were introducing facets, shows that ideas and concepts in architectural and building sciences are tangible enough to allow for an almost immediate definition of facets relevant for that domain of knowledge. The proximity of ideas on architectural and building sciences to human senses make the definition of facets more intuitive.

Unfortunately, the definition of ideas and concepts in financial sciences is mostly non-spatial, and as a consequence, the designation of facets in our specific case not as intuitive. In financial sciences, a researcher would be more interested, for example, in the way changes in quantitative measurements, over discrete time, would affect the price. These are abstract concepts, and as a consequence, it is hard to describe them through concrete, tangible similarities.

According to the representational process defined in Section II facets are defined based on two inputs: intrinsic requirements of a domain of knowledge and a computational taxonomy. The requirements for a computational representation for economics were previously defined in Section IV. A computational taxonomy, as previously defined in Section II.A, is an inventory of computer technologies available and relevant for the implementation of domain-specific cases. The specific computational taxonomy in use is explored during the definition of each facet when we examine technology alternatives.

---

[11] Examples are a sequence of prices, in which a specific algorithm tracks features of price variations over different time windows, e.g., during the last hour, a day, a week. Different windows can be compared with adjacent or non-adjacent windows for identification of useful patterns.

Over the next sections, we detail the exercise to find out the relevant set of facets for our domain of knowledge: economics. For that, we formalize the four facets required for the definition of aspects that make up subjects and ideas in the field of economics: streaming, reactives, distribution, and simulation.

### A. Streaming

The original idea of streams, put merely, starts with a vision of a graph in which nodes are processors and edges are communication paths. Each node holds incoming and outgoing communication paths to other nodes in the graph. The basic idea of streaming relates to continuous sequences of data fragments traveling over communication paths, in which each node executes specific tasks upon arrival of fragments of data.

Streams are traditionally used in domains where concurrency and speed of processing is a core requirement. Some of those domains include micro-hardware control, image processing, graphics, sound processing, compression, networking, encryption, and digital signal filtering [30] [31]. Given similar requirements around performance, time series, sliding time windows computations, and the graph-oriented nature of financial models, listed previously in Section IV, streaming is selected as the first facet in a computational representation for the field of economics.

#### A.1. Models of Computation

Streams have been used as a notation for representation of computational elements in domains of knowledge outside of economics for a long time [32]. The first reference to an equivalent paradigm was on bullet notes given by Douglas McIlroy [33] on October 11th of 1964.

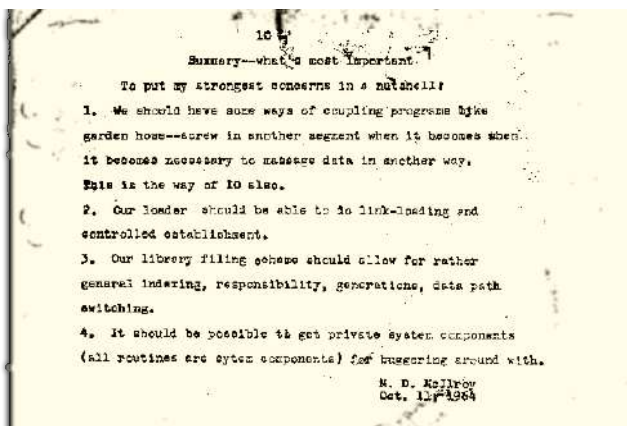

**Figure 3. The First Reference to Streams**

A bullet summary by Douglas McIlroy on "what's most important", suggesting a function to "have some ways of coupling programs like garden hoses", what was referred by subsequent literature as *streams*

The original insight of "digital hoses", coined by Douglas McIlroy [33] evolved through different milestones to consolidate the idea of streaming systems [32] [31] [34]. Each milestone of the evolution of what were initially *digital hoses* refers to a specific model of computation, as shown in Figure 4.

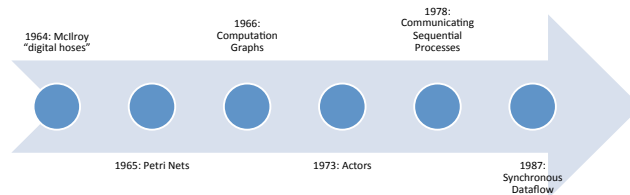

**Figure 4. Timeline of Evolution: Models of Computation**

Over time the original idea of "digital hoses" evolved on variations called "models of computation" from Petri Nets, Computation Graphs, Communicating Sequential Processes, and Synchronous Dataflow.

Each of these milestones, or models of computation, present different features and define a computational taxonomy of streaming systems [31]:

- **Petri Nets**[12]: a directed bipartite graph, where nodes either represent transitions or conditions. A directed edge specify which pre or post conditions are a requirement for a transition [35] [36] [37] [38];

- **Computation Graphs**: a graph-theoretic model for the description and analysis of parallel computations where computation steps correspond to nodes of a graph, while branches represent a dependency between computation-steps. Each branch is associated with independent queues of data [39];

- **Kahn Process Networks**: a distributed model of computation where deterministic sequential processes are nodes, and FIFO channels are the edges of a graph network [40];

- **Actors**: a graph-based model of concurrent computation in which nodes are actors, and upon receipt of messages an actor can send new messages or create new actors. In this sense, edges can be created on demand and indicate a communication by message-passing [41] [42] [43] [44] ;

- **Communicating Sequential Processes**: a textual and formal language for describing concurrent interaction

---

[12] Despite of Petri's original thesis of 1962 [35] the formalization of Petri Nets as they are currently known only came a bit later, in a 1965 colloquium [36], published in 1967 [37].

based on primitive processes and events. Primitive processes are fundamental behaviors, and events represent indivisible and instantaneous interactions [45] [46];

- **Synchronous Dataflow**: a particular case of data flow in which each node represents a function, and each arc represents a signal path. It is a simplification of Kahn Process Network by limiting the number of messages each node consume and produce per signal [47].

Despite lacking a standard nomenclature, topology, or modes of communication, all models of computation of streaming systems can be represented on a higher level by nodes and edges, arranged as graphs. In fact, the specific features of the models of computation can be normalized over three specific features [32] [31] [34]: topology, determinism, and dynamicity.

- **Topology**: defines the way in which nodes are arranged in a network;
- **Determinism**: establishes if the final results of execution are always the same, given the same set of inputs;
- **Dynamicity**: establishes if execution parameters (i.e., amount of buffering and communication patterns) can be decided and arranged statically and dynamically (i.e., at compilation time or runtime) [30] [34] [31] [48]

The streaming facet translates these normalized features of models of computation – topology, determinism, and dynamicity – into three specific properties of financial models: synchronicity, connectivity, and plasticity. These properties are explained over the next topic when we explain the mechanics to define financial models using streams.

*A.2. Defining Financial Models as Streams*

The streaming facet defines a graph-oriented domain-specific language [49] [50] to define financial models as a route of fragments of meta-data $x$ through a chain of reusable and exchangeable processors $P_i$. The chain of processors $P_i$ is arranged as a function composition, as described in Equation 1 [51] [52] [53].

$$(P_1 \; o \; P_2 \; o \; ... \; o \; P_n)(x)$$

**Equation 1. Function Composition**

In the specific representation for the domain of economics, processors are chained together by a synchronicity operator $\delta$ giving a composition of processors the form shown in Equation 2.

$$x \rightarrow P_1 \; \delta_1 \; P_2 \; \delta_2 \; ... \; \delta_{n-1} \; P_n$$

**Equation 2. Composition by Synchronicity Operator**

We call this chain of processors $P_1 \; ... \; P_n$ connected by $\delta$ a stream. Equation 3 gives an equivalent graph representation, based on edges and vertices, of the same stream.

$$\phi = (P_i, \delta_i)$$

**Equation 3. Graph-Oriented Representation of a Stream**

In Equation 3, $\phi$ is a directed sub-graph $\phi(V, E)$ in which $V$, the set of vertices $P_i$, are processors, and E, the set of edges $\delta_i$, are synchronicity operators. The same graph $\phi$ can be visualized as a connected directional graph, as shown in Figure 5.

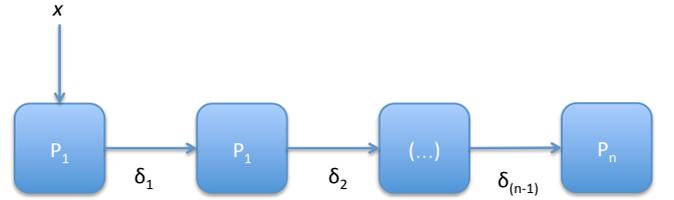

**Figure 5. Streams as a directed graph**

A stream can be visualized as a connected graph $\phi(V, E)$, in which edges are given by synchronicity operators $\delta_i$ and vertices, or nodes, by processors $P_i$

We assume that a financial model, to be defined in terms of requirements listed in Section IV, must carry three fundamental properties:

- **Synchronicity**: financial models must operate on data fragments $x$ synchronously or asynchronously, where $x$ is defined in Equation 1;
- **Connectivity**: financial models are created by the composition of smaller, modular pieces that can often be recursively leveraged as smaller, reusable models;
- **Plasticity**: the composition and the behavior of a financial model can change, in real-time, upon arrival of new data fragments $x$, as defined in Equation 1.

Each of these three fundamental properties – synchronicity, connectivity, and plasticity - is formalized as stream elements, or interchangeably [13] as graph properties. Over the next sections, we formalize the representation of financial models over a stream-oriented language based on these three fundamental properties.

*A.2.1. Synchronicity*

In a stream $\phi$, as described in Equation 3, processors $P_i$ spawn tasks $t_{s,i}$ in pools of tasks $T_i$ to handle data fragments $x$ as they arrive. Each pool holds a variable number of tasks $s$, where the exact value of $s$, is irrelevant and associated with scheduling configuration details.

The synchronicity operator $\delta_i$ indicates how a fragment $x$ is "handed over" from tasks in pools $T_i$ in processors $P_i$, to $P_{i+1}$, where each $\delta_i$ can indicate two distinct modes: synchronous and asynchronous.

---

[13] As we have shown in Section V.A.1 (through the different models of computation of streams) graph or stream representations are functionally interchangeable [32] [31] [34].

In synchronous mode, $T_i$ depends on the completion of $T_{i+1}$, and therefore $T_i$ can only proceed, and consume the next fragment $x$, after termination of task $T_{i+1}$.

Alternatively, in asynchronous mode, $T_i$ does not depend on the completion of $T_{i+1}$, and therefore $T_i$ can consume the next fragment $x$ regardless of the result and termination of $T_{i+1}$.

*A.2.2. Connectivity*

Financial models are represented as directed graphs composed of a limited set of directed sub-graphs $\phi$, as described in Equation 3, bound together by connectors. For all purposes, a connector $C$ is a specialization type of processor $P$, as defined in Equation 1 as $P_{1..n}$, so that $C \cong P$.

As a specialized processor, a connector carries additional properties to allow the connection of multiple streaming sub-graphs $\phi = (P_i, \delta_i)$ into larger, interconnected networks of streams.

The composition of more elaborate, interconnected networks of streams allows the support of more complex functions. These functions include the plasticity property, described through a graph modification connector in the next section, reactive behaviors described in Section B, the distribution facet described in Section C, and enabling of distribution spaces described in Section C.2. A complete outline of possible connectivity functions is provided later in this paper, in Section VI.B, when we describe in details the processor contribution.

*A.2.3. Plasticity*

A graph representing a financial model should be able to modify itself upon the arrival of relevant data fragments, depending on specific requirements of the model under study.

In the scope of this research, the ability to modify a graph $\Phi$ on demand is referred to as plasticity and is given by a special modification connector $C_p$. The connector $C_p$ is given by function $f$ of a predicate $P$ on data fragment $X$, and a sub-graph template $\bar{\phi}$, formalized by Equation 4.

$$\bar{\phi} = (\bar{P}_i, \bar{\delta}_i)$$
$$P: X \to \{true, false\}$$
$$C_p = f(P, \Phi, \bar{\phi})$$

**Equation 4. Definition of Plasticity Function**

Plasticity occurs upon arrival of data fragment $X$. In case $P$ resolves as a $true$ for $X$, a sub-graph $\phi$ based on template $\bar{\phi}$ is appended to graph $\Phi$.

In short, this model in Equation 4 allows a graph to modify itself if an arrival of a data fragment $X$ causes the predicate $P$ to resolve as $true$.

The best way to explain the plasticity property is through an example, and preferably in finance, through a common use case. In Figure 6, we depict the example of a definition of a new pricing route for a stock that should be set up upon arrival of a new symbol of that stock on a sequential feed of price ticks.

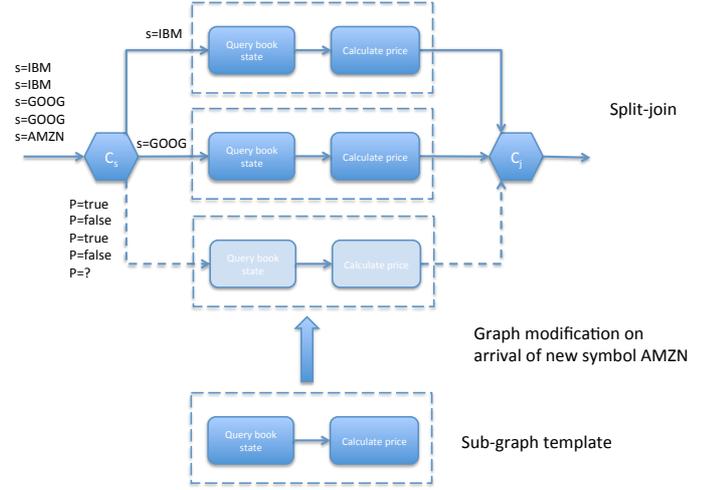

**Figure 6. Graph Modification Connector Example**

An example of plasticity through the application of a graph modification connector, when new symbols arrive and predicate $P$ fires $true$, a new path on the graph is added based on the sub-graph template. In this example, the arrival of symbol AMZN will create a new branch and the modification of the overall graph.

On this use case it is assumed high-frequency requirements, so to maximize throughput every branch on the graph is dedicated to one symbol. The complete set of symbols is not known in advance. Therefore, a new branch must be created for every new incoming symbol, the first time an instance of this symbol is received. On this exercise, each branch executes the following steps for any given price tick:

- Query current state of the order book for current bid and ask prices of stock $s$;
- Calculate the price of a stock based on current mid-price, spread, and exponentially weighted moving average of the mid-price[14].

Every incoming fragment $x$, arriving at time $t$, carries a tuple $(s, p)$ where $s$ is symbol and $p$ is the price. In this exercise, for the sake of simplification, since our concern is specifically to exemplify the property of plasticity, we are only interested in symbols. The sequence of incoming symbols is given on this example by the sequence in Equation 5.

$$s: (IBM, IBM, GOOG, GOOG, AMZN, ...)$$

**Equation 5. Sequence of Incoming Symbols**

In response to each item in the sequence, predicate $P$ turns to $true$ if this is the first arrival of that symbol on the sequence. In response to the sequence of symbols in Equation

---
[14] An example of signal attenuation functions are filters based on exponentially weighted moving average processors [8].

5, $P$ yields a correspondent sequence of results given by Equation 6.

$$P: (true, false, true, false, true, \ldots)$$

Equation 6. Sequence of Predicate Results

In the previous Figure 6, the configuration of the graph $\Phi$ is shown as a snapshot in time right after the arrival of the AMZN symbol, as result of the use of a modification connector $C_p$. Before that snapshot, the symbol IBM had arrived, producing sub-graph $s = IBM$, followed by symbol GOOG, what produced the sub-graph $s = GOOG$.

On arrival of symbol AMZM, the predicate $P$ yields $true$ and the third branch is created and associated with the newly arrived symbol AMZN. From that point on, arrivals of new symbols AMZN will be routed through the newly created branch.

The plasticity property allows for on-demand modifications on connections of a graph representing a financial model. This example is an important and common use case on models related to the trading of financial instruments.

### B. Reactives

As stated on requirements defined in Section IV, financial models must maintain continuous interaction with an ever-changing state that varies over time and is external to the financial model at hand. Rules on the financial model have to trigger specific actions based on external events that can occur at unpredictable times.

These external changes are by nature unpredictable, and by definition are hard to represent in conventional, sequential programming. External changes are associated with events, and require a number of non-sequential[15] properties for representation in financial models: inverted control, abstraction of time management, and abstraction of synchronicity details [54]:

- **Inverted Control**: Financial models keep a continuous and persistent interaction with their execution environment, executing actions based on events triggered by external sources. External sources then drive the order of execution, and as a consequence, in many particular cases, the rules and control flow of a financial model is inverted [54].

- **Abstraction of Time Management**: Financial models often require a notion of a discrete time series, in which the modification of the event associated with each time $t_i$ in the time series is performed by behaviors [55] [54]. The event is associated to either a lifecycle change (e.g., corporate actions in a stock, roll-over operations in derivative instruments) or variations of value (e.g., the price of an asset, ratio of risk exposure) over time. After a relationship between reactive entities is set, computation dependencies and handling of events over time are automatic and the representation of time is intrinsic to every event [56].

- **Abstraction of Synchronicity Details**: Financial models require the abstraction of synchronicity details in the event-driven communication. Financial entities are often defined in terms of relationships with other entities. In a representation suitable for financial models, associations are established declaratively, similar to the way in which cells in a spreadsheet are defined and associated with a formula[16]. The declarative association through formula provides automatic management of associations between data dependencies. The event-driven communication synchronizing the state of those entities is intrinsic to the representation of the association and therefore transparent.

Functionally, these properties – inverted control, abstraction of time management and details of event-driven communication - are related in computer science to what is commonly called reactive programming [55], and referred in the scope of this research as a reactive[17] facet.

The *reactive facet* is a declarative paradigm that allows the definition of what has to be done through reactive relationships, and let the computational representation automatically take care of when to do it, and who gets affected. A similar and more intuitive model is exemplified by a number of cells in an electronic spreadsheet representing a formula. Similarly, reactives allow for an intuitive representation of primitives and formula, in which composition of formula from primitives and other formula is defined declaratively [57] [54].

To describe declarative associations of reactive variables, we take for example the simple formula in Equation 7.

$$A = B + C$$

Equation 7. Reactive Formula Example

In a sequential representation, variables $B$ and $C$ would have to be set first, so that only then the computation of $A$ could occur. Alternatively, in a reactive representation, the formula is declared first, setting a graph of reactive dependencies. In Figure 7 we show the graph of dependencies for the formula in Equation 7.

---

[15] Some literature considers reactives an extension of stream processing [32]. Given the nature and requirements of financial models we opted to differentiate between sequential (streams) and non-sequential (reactives) as two separate and yet complementary facets.

[16] The designation of a formula is equivalent to the concept of a formula in an electronic spreadsheet

[17] Functionally equivalent patterns like observers, event-driven programming and asynchronous callbacks were also considered as possible alternatives to reactives, but unfortunately they carry their own impeding limitations. The coordination of individual callbacks, over a shared state, across numerous code fragments, in which the order of execution cannot be predicted, is an error prone, cryptic, daunting programming task [54]. Additionally, since callbacks do not produce a return value, these alternative programming patterns must perform side effects in order to affect the application state [79].

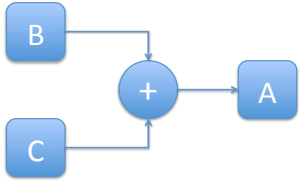

**Figure 7. Graph of Reactive Dependencies**

The reactive graph, representing a simple formula $A = B + C$. A formula of functions, operators, or other reactives is set as a graph of communication between reactives.

The graph in Figure 7 represents that, in case of a change on the value of either $B$ or $C$, the executing environment abstracts the notion of a discrete time change and event-driven communication by propagating the modification across all dependencies in the graph. The exact way a value propagates over time through a graph of dependencies, in this case from $B$ or $C$ to $A$, can occur in more than one way and is abstracted from the representation itself.

The reactive paradigm is a broad concept, subject to a specific classification in terms of basic features, evaluation model, lifting, and directionality. These classes are related to special considerations for the use of the reactive paradigm is used to represent financial models, according to requirements defined in Section IV.

*B.1. Basic Features*

Two basic features define the reactive programming paradigm: behaviors and events. They are often referred to as *duals* because one can be used to represent the other. The behavior feature refers to mutable, time-varying values. The event feature refers to potentially infinite, immutable modifications that occur at discrete points in time.

In the computational representation for the field of economics, behaviors are associated with specialized processors $R$ so that $R \cong P$. The dual of $R$ is a virtually infinite sequence of events $x$, as previously discussed in Section A.

Given for example two disjoint sub-graphs $\phi'$ and $\phi''$, a generic synchronicity operator $\delta$, and behaviors associated to reactive processors $R'$ and $R''$, as described in Equation 8.

$$\phi' \; \delta \; R'$$
$$\phi'' \; \delta \; R''$$

**Equation 8. Composition of Streams and Reactives**

A reactive function $f_r'(R', R'')$ is evaluated on the arrival of either $x'$ or $x''$, in each of the streams defined by sub-graphs $\phi'$ and $\phi''$, as shown in Figure 8.

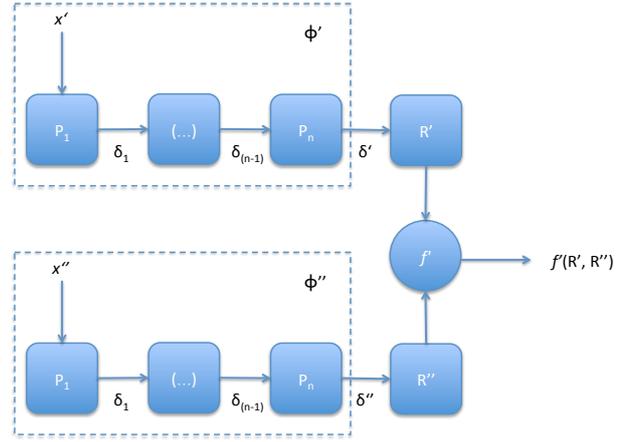

**Figure 8. Composition of Streams and Reactives**

Reactives are specialized processors in a stream, either as reactive steps in linear streams, like $R'$ and $R''$, or as connectors for disjoint sub-graphs, like the reactive function $f_r'$.

As represented in Figure 8, reactives are specialized processors in a stream. In the case of $R'$ and $R''$, they have to behave like regular, sequential processors for sub-graphs $\phi'$ and $\phi''$, and at the same time act as reactives for the reactive function $f_r'(R', R'')$.

Reactive dependencies and respective flow of execution are defined declaratively. As a consequence, after the reactive graph is defined, calculations are not affected by the sequence of initialization of $R'$ and $R''$.

*B.2. Evaluation Model*

The evaluation model of the reactive facet defines how a change $x$ in stream $\phi$ propagates through a dependency graph of values and computations.

In a pull-based evaluation model, a value is calculated on demand, or in other words, a value has to be "pulled" from the source whenever required.

On the other way, in a push-based evaluation model, every change in value has to be sent to dependent computations. The push-based propagation is called data-driven since it occurs by the availability of new data.

The evaluation model has no direct implications on the representation, but it does, however, have implications on the distribution facet. Those implications are discussed in detail in the upcoming Section V.C, related to the distribution facet.

*B.3. Lifting*

We call lifting the transformation of a generic function $f(x)$ applied to $x$ to a lifted function $f'(R<x>)$, where $R<x>$ is a reactive, or behavior, type of $x$ [54] given in Equation 9.

$$lift: f(x) \rightarrow f'(R<x>)$$

**Equation 9. Reactive Lift Function**

When looking at time step $i$, the resolution of a lifted function $f'$ on value $x_i$ yields the original function $f$, as shown in Equation 10. Mathematical operators (e.g., $+, -, *$) and user-defined functions, respectively, are functionally equivalent to lifted operators and functions.

$$f'(R<x_i>) \rightarrow f(x_i)$$

**Equation 10. Original Lifted Function**

The representation of the lifting transformation is classified further in terms of how much additional context is needed whenever an operator or a function has to be lifted to a reactive representation. This classification defines a lifting transformation as manual, explicit, or implicit.

- **Manual**: on manual lifting, a representation does not provide transparent lifting. A time-varying value has to be manually extracted and applied to operators, functions, or dependent variables.
- **Explicit**: on explicit lifting, the representation defines a number of unique operators that can be used to lift a function $f$ to $f'$.
- **Implicit**: on implicit lifting, all operators and functions of a representation applied to $x$, user-defined or not, are transparently lifted to a reactive item $R<x>$.

For simplicity of communication, in the computational representation for the field of economics, all reactive transformations are implicitly lifted. This requirement will impose additional constraints on candidate implementations, but as a consequence gives a higher level of abstraction to the representation.

*B.4. Directionality*

A reactive representation may allow reactive propagation of changes to occur in one direction – unidirectional – or in either direction – multidirectional. In requirements listed in Section IV, there were no specific cases in which multidirectional propagation was an absolute requirement. As a consequence, for simplicity, for a computational representation for the field of economics, only a unidirectional propagation is required.

C. *Distribution*

In Section IV we listed some requirements explicitly related to the operation of financial models in large scale, both in terms of computational power and storage.

Of those requirements, two specifically - virtually infinite historical records, and responsiveness – require the use of distributed resources [58] to be able to scale to more than a single processing or storage unit. The *distribution facet* gives the computational representation the ability to communicate functions related to scaling up the workload of a financial model across multiple processors.

A distribution facet is, in essence, a particular application of connectors, as described in Section A.2. A connector $C$ is a specialized type of processor $P$ so that $C \cong P$. That specialization means that in addition to the behavior of processors, a connector carries additional properties to allow the composition of streaming sub-graphs $\phi = (P_i, \delta_i)$ into larger, interconnected networks of streams.

On its more generic form, any connector $C'$ allows for bridging of a number $n$ of incoming sub-graphs $\phi'$ and a number $m$ of outgoing sub-graphs $\phi''$, as described in Figure 9:

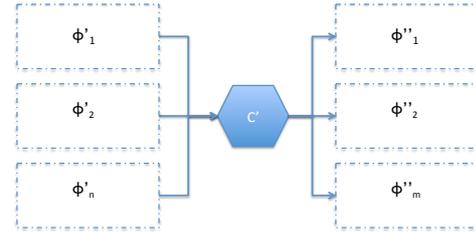

**Figure 9. Connectors and Incoming and Outgoing Streams**

Connectors in the distribution facet are used to compose streams by bridging a number n of incoming sub-graphs $\phi'$ and m outgoing sub-graphs to build more elaborate graphs.

The generic description of the distribution facet as connectors for a $n:m$ association of incoming to outgoing sub-graphs of streams has two significant consequences: improved expressiveness of the streaming notation, and leveraging of distributed and parallel processing in large scale.

*C.1. Improved Expressiveness*

The first consequence, improved expressiveness, is related to the possibility of laying out streams and connectors in different combinations to define more elaborated patterns [50]. The placement of connectors in different locations of the streaming graph can define structures like split-joins and feedbacks [30] [34] [31] [48] [32] [59]. A visual representation of split-join and feedback loop is given in Figure 10.

A split-join pattern is given by a pair of connectors, $C_s$ and $C_j$, positioned around a set of processors. The connector $C_s$ is placed on the splitting, inbound edge of the set of processors, while the connector $C_j$ is placed on the joiner, outbound edge of the set of the processor.

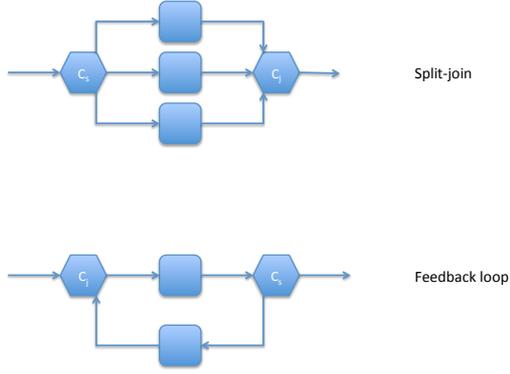

**Figure 10. Connectors and Communication Patterns**

Combination of connectors used to create more elaborate composite patterns like split-joins and feedback loops.

Edges for communication in and out of the processor stack between $C_s$ and $C_j$ are given by the asynchronous operator $\delta$. As a consequence, fragments leaving $C_s$ might hit all processors concurrently, and the order of execution of these processors cannot be guaranteed. Unless specialized processors are inserted in the flow before the join of $C_j$, with the ability of re-establishing the original order of execution, the overall execution is non-deterministic.

A variation of the split-join pattern, the feedback loop, given in the lower part of Figure 10, is a re-arrangement of $C_s$ and $C_j$ to represent a loopback of data fragments. Edges for communication out of $C_j$ and into $C_s$ are asynchronous, i.e., the synchronicity operator $\delta$ is of type asynchronous.

Since there is a requirement of asynchronous communication on edges from and to $C_s$ and $C_j$, results of the execution of the overall stream in a feedback loop are non-deterministic.

*C.2. Parallelism and Distribution Spaces*

The second consequence is the possibility of describing an execution flow spanning multiple computational environments and locations concurrently. Each of those disjoint computational environments is called *space*. A space by definition can be associated with different processors, in different locations, as required.

For example, given a connector $C'$ and disjoint sub-graphs $\phi'$, $\phi''$ and $\phi'''$ on a specific composition, as described in Equation 11:

$$\begin{array}{l}\phi' \; \delta \; C' \\ C' \; \delta \; \phi'' \\ C' \; \delta \; \phi'''\end{array}$$

**Equation 11. Connectors and Distribution Spaces**

Following the definition of the connectivity property of financial models, described in Section A.2, a larger graph $\Phi$ is defined as a result of connector $C'$ applied on sub-graphs $\phi'$, $\phi''$ and $\phi'''$ as shown in Figure 11.

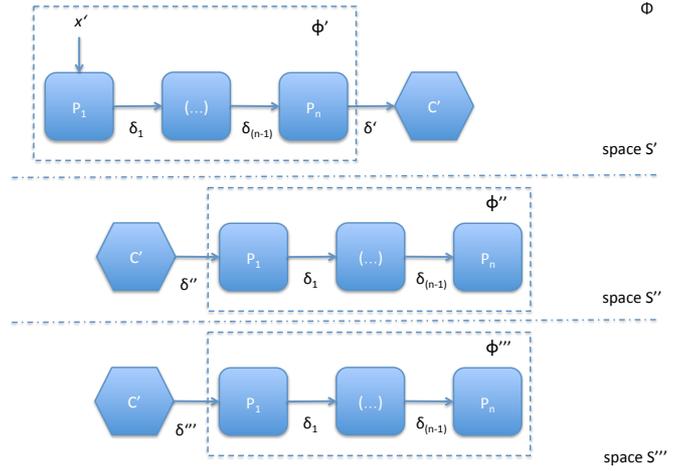

**Figure 11. Graph Composition by Connectors and Spaces**

The application of connectors to sub-graphs defines a generic graph $\Phi$ and multiple spaces $S'$, $S''$ and $S'''$, which can be optionally associated to computational resources spanning multiple locations.

Each space is a fragment of a complete graph $\Phi$, in which boundaries of any space are given by incoming or outgoing edges of a connector. Each space abstracts details related to distribution or concurrency aspects of a financial model and can be at a later time associated to computational resources spanning multiple locations, without affecting the high-level representation of the financial model itself.

A complete graph $\Phi$ in Figure 11 shows the use of a connector $C'$ to define multiple spaces $S'$, $S''$ and $S'''$, each associated to sub-graphs $\phi'$, $\phi''$ and $\phi'''$. Each space and sub-graph can be associated to different distribution contexts, without affecting the intuitive description of a financial model.

In essence, connectors allow scaling of a financial model to handle a virtually infinite load and volume of data by adding the notion of locality and distribution transparently through the use of spaces.

This notion is intrinsic to the representation in a sense that it is not defined in the financial model described, and decisions relative to performance, storage, and processing power can be made at a later time, with no modifications to the financial model itself.

*D. Simulation*

Financial models are a representation of complex systems in which the intent of defining one, in many cases, is to allow prediction of outcomes, through the application of disciplined, objective research methods.

Simulations are a fundamental technique for research of complex problems in many disciplines, especially in financial sciences, through the application of specialized algorithms [60] to define, search and test possible viable solutions. Simulations have an exact placement in a proof pipeline for crowd-based investigation in economics [5], or in other words, simulations are the imitation of a system [61]. Financial models are, in essence, an imitation – a controlled simplification to the right scale – of large, complex systems.

The facet simulation is responsible for representing methods allowing the anticipation of possible outcomes in financial models. On that regard, the general topic of simulations is extensive and under active research [5]. The primary challenge then, when defining a simulation facet in a domain of knowledge is to establish the exact level of simplification that can be applied to models on that representation, without affecting the quality of insights into the central problem under simulation.

According to the representational process previously defined in Section II, any facet, and in this case precisely the simulation facet, must be selected based on domain-specific requirements and a computational taxonomy. Domain-specific requirements were previously defined in Section IV, and given the importance of the subject of simulation, should be augmented by the proof pipeline for large-scale collaboration in economics [5] [6]. A computational taxonomy is given by various alternatives for classification simulation techniques [62] [61] [63] [60], shown in Figure 12. According to that representational process, the combination of requirements and techniques are enough to select and adjust relevant properties of simulation for financial models.

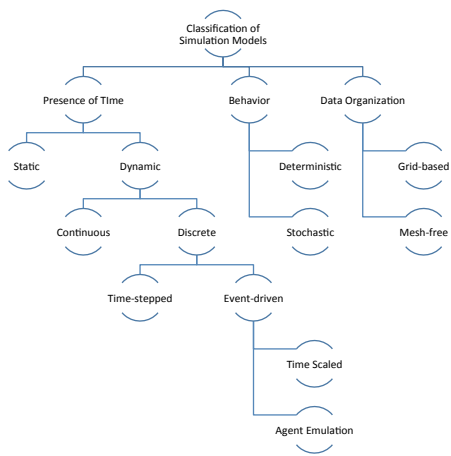

**Figure 12. Simulation Taxonomy and Relevance to Economics**

Simulation facet classified according to a generic taxonomy based on the presence of time, behavior and data organization. Leafs marked in a dotted like do not represent relevance to the field of economics.

The taxonomy of a simulation facet organizes all possible representations in three dimensions according to the presence of time, behavior and data organization. All three dimensions of classification are complete and not mutually exclusive, in a sense that a model requires a concomitant classification in each of the dimensions.

For example, a model that anticipates the effect of corporate actions over the price of an asset is static, deterministic and grid-based, while a model that uses random shocks to determine the influence of multiple features in the profitability of a portfolio in closing prices is dynamic time-stepped, stochastic and grid-based.

*D.1. Presence of Time*

The first classification takes into consideration the presence of time [63], or time of change [60]. As the name implies, this classification considers if time is a significant variable in defining the outcome of a simulation [62].

Under this classification, system models can be classified as static or dynamic, respectively by observing if the system can be adequately modeled without or with a variable associated with time.

In our cases of use, previously defined in Section IV, it is clear that the absolute majority of financial models are dynamic. However, we cannot ignore that some critical exceptions do not require the presence of time. An example of a static system relevant to the field of financial sciences would be the influence of corporate actions on the valuation of equity assets on the day that a specific action took place.

Dynamic systems are further classified as either continuous or discrete [63]. Continuous dynamic systems consider that variables or features into consideration evolve continuously and are usually subject to modeling through differential equations, representing continuous modifications of a system. Some examples outside of economics are often related to classic mechanics like particles moving in gravitational fields, or an oscillating pendulum [62]. All observable phenomena are by nature continuous, but since by definition models are surrogates of real events, the use of discrete dynamic systems allow a significant simplification by considering that all variables of the system are piecewise constant functions of time, only possessing one of many values within a finite range.

Dynamic discrete systems can be classified even further depending on the irregularity of the time interval as time-stepped or event-driven systems.

On time-stepped systems, time intervals are constant or derived from fractions of time in which periodicity can be clearly ascertained. Examples of a dynamic discrete system in finance are models associated with changes in discrete values (e.g., prices, returns, risk ratios) over a time-series. Dynamic discrete time-stepped systems account for the majority of the cases of use in finance.

On event-driven systems[18], time interruptions occur in irregular intervals, driven by external sources of the model itself. In finance, such systems are not as usual as systems based on constant time steps. Event driven-systems should, however, be considered at least as necessary, and an adequate tool when investigating sophisticated use cases. Some examples are related to cases of agent-based simulation of a central limit order book. In these simulations, software agents play the role of market participants and are used to gauge the influence of real-world economic agents to study the effect of a pre-defined behavior in variations of the price of financial instruments [24] [25] [26] [27] [28] [29].

Dynamic discrete systems are represented in a computational representation for the field of economics by adjusting generic streams $\phi(P, \delta)$ in two specific points to represent either time-stepped systems or event-driven systems.

Time-stepped systems are replicated by replacing the generic endpoint $X$ of fragments $x$ in a stream $\phi(P, \delta)$ by a time-paced endpoint $f(T, X)$ so that the period $T$ between events $x_t$ can be adjusted, as shown in Figure 13.

---

[18] We assume agent-based modeling is an extension of tools commonly used to simulate event-driven dynamic systems [82] [81].

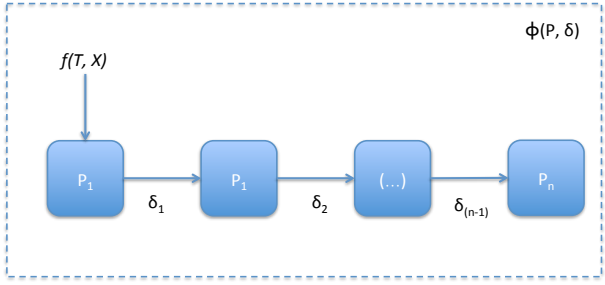

**Figure 13. Simulation of Time-Stepped System**

A time-stepped system is simulated by replacing the generic endpoint $X$ of fragments $x$ by a periodic function $f(T, X)$ in which the period $T$ is a fraction of the time-step present in the original system under simulation.

Adjusting $T$ to minimal amounts allows for the simulation in milliseconds of market behaviors that otherwise could only be observed over long periods, achieving for all practical purposes a time-squeezing effect. The simulation can then be replayed as many times as required, with different values for all relevant features [8].

Event-driven systems are represented in a computational representation for the field of economics by two different variations: time scaling and agent emulation.

In the time scaling variation, a generic endpoint $X$ of fragments $x$ in a stream $\phi(P, \delta)$ is replaced by an endpoint $f(k, X)$, allowing event-driven systems to be replicated by replaying events $x_t$ from $X$ in a different scale of time. The same time-squeezing effect observed in time-stepped simulations is achieved by adjusting the time of occurrence of each event $x_t$ to a shorter scale $k$ as $x_{t'}$ as described in Equation 12.

$$t' = t_0 + \frac{1}{k}t \qquad \text{Equation 12. Time Scaling}$$

Where $t_0$ is an arbitrary time assigned to the beginning of the simulation, and $k$ is the time compression scale.

In the agent emulation variation, shown in Figure 14, streams $\phi(P, \delta)$ play the role of individual software agents, similar to what some literary references call a process-oriented paradigm [64], process-modeling [65] or process-interaction [66].

Each software agent $A_i$, represented by stream $\phi(P, \delta)$, is used to model real-world entities that hold state and evolve in time. Agents interact through a shared context, either by direct communication or by modification of state in a shared resource.

A component called a Discrete-Event Simulation Environment is responsible for proper scheduling and coordination of an agent $A_i$ by issuing and capturing variations of events of type activate, cancel, or yield.

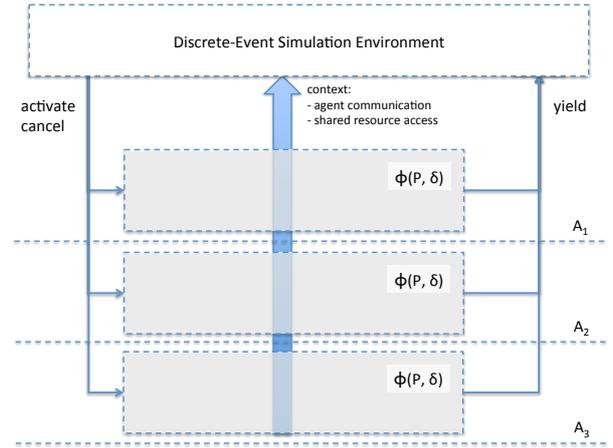

**Figure 14. Discrete-Event Simulation Environment**

Each stream $\phi(P, \delta)$ plays the role of software agents $A_i$, and execution is done by specific signals activate, cancel or yield. Agent communication and competing access for shared resources is done through the agent context.

In a higher level, an activate signal marks an agent as eligible for execution, while a cancel event forces an agent to release resources and yield execution. Agents notify the discrete-event simulation environment of specific changes in a task status by issuing different types of yield signals. A yield signal tells the discrete-event simulation environment that the agent can be de-scheduled and action can go to an eligible agent if such an agent is available [64] [67].

*D.2. Behavior*

The second classification of simulation models takes into consideration the randomness of results of the execution of a model given a constant set of inputs.

In deterministic models, the result of a model execution depends only on the input given to the model, what means that repeating a simulation several times will yield the same results [63]. On the other hand, in stochastic models, the result of a simulation varies randomly[19] [68].

Both deterministic and stochastic behaviors are required for financial models, and the exact nature of a model is defined by behaviors of processors and topology of the underlying graph $\phi$ describing the model, as explained previously in Sections A, B, and C.

*D.3. Data Organization*

The third classification of simulation models arranges simulations as grid-based or mesh-free, depending on the data organization scheme [60].

In the mesh-free organization [69], data is associated with individual and disconnected (i.e., mesh-free) nodes called particles. Updates to a particle are not bound to neighboring or connected relationships between particles, but instead are related to interactions to all particles considered relevant. The mesh-free organization enables the simulation of complex systems, at the expense of computing power and programming complexity. Use of mesh-free simulation in finance usually applies to overspecialized cases of use [70] [71] [72] [73] and as a result was considered out of scope and left out of the list of cases of use in Section IV.

---

[19] Pseudo-random models, in which a random outcome is emulated by a pre-defined sequence of random values, based on a number called a *seed*, are indeed a special case of a deterministic model.

Alternatively, in the grid-based organization, the state of a simulation is arranged in discrete cells at particular locations in a grid. Updates occur to each cell based on previous state and those of its neighbors, or to cells to which it is connected. The absolute majority of the financial models are grid-based.

In this proposed computational representation for the field of economics some of the fundamental constructions - reactive primitives, functions, and operators - play the role of cells in a grid-based organization while connections reactive primitives constructions are arranged in the same way as cell dependencies.

## VI. CONTRIBUTIONS

As defined by the representational process introduced in Section II, the second element of a computational representation is referred to as contributions. As introduced in Section II.B, contributions are defined as shareable and formal evidence of a scientific crowd-based investigation.

According to the representational process in Section II contributions are a taxonomy of shareable evidence that is relevant to cases of use on the domain of knowledge under study, in this specific case, economics. The cases of use were described previously in Section IV. According to the evidential properties discussed in Section II.B, all contributions must follow a classification system, called taxonomy of contributions. Contributions for a computational representation for the field of economics must cover a broad range of models, methods, and results relevant to financial sciences [74]. Some examples include datasets in small, medium or large scale; time series in low, medium or high frequency; calculation processors and visualization plots; and results related to historical and real-time execution, simulation and backtesting. The taxonomy of contributions for the field of economics is shown in Figure 15.

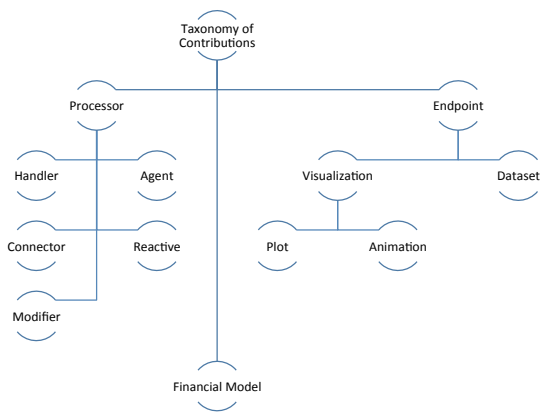

**Figure 15. Taxonomy of Contributions**

Contributions are classified as financial models, processors, or endpoints. Endpoints are either visualization (i.e., plots, animations) or datasets.

All contributions carry the evidential properties defined in Section II.B, and therefore, in addition to falling in precisely one classification of the tree in Figure 15, all contributions are also uniquely identified, carry a detailed record of provenance and hold enforceable ownership and access information. On a higher-level, contributions are classified as financial models, processors or endpoints.

### A. Financial Model

The first type of contribution for a computational representation for the field of economics is a financial model. Financial Models are by definition a description of observable phenomena in the field of economics, simplified to the right scale, and adjusted for use in the process of a crowd-based investigation [5]. An extensive outline of requirements of financial models is listed in Section IV.

Since financial models are contributions, they follow what we call evidential properties of contributions, explained in Section II.B. As such, financial models can be defined and reused by different users.

From a representational perspective, financial models are built based on streams of processors and endpoints, arranged as graphs. The fundamentals for the description of financial models as streams are described in Section V.A.2 on page 8. An example of a generic financial model is depicted as graph $\Phi$ in Figure 11 on page 13. Processors and endpoints as contributions are explained over the following sections B and C respectively.

### B. Processors

The second type of contribution for a computational representation for the field of economics is a processor. Processors are steps on the execution stream and are placed to perform specific computations on fragments of data $x$, as already explained in details in Section V.A.2.

Since processors are contributions, they carry evidential properties of contributions, explained in Section II.B. As such, processors can be defined and reused across different financial models (i.e., execution streams, as explained in Section V.A.2), by different users, whenever that same specific function is required. Processors are further classified as handlers, connectors, modifiers, reactives, or agents.

- **Handler**: the simplest type of a processor is a handler. A handler applies transformations to a fragment of meta-data $x$ as defined in Section V.A.2;

- **Connector**: the composition of larger, more complex financial models from multiple smaller sub-graphs is possible by using a specialized type of processor called connectors. Connectors were explained in details in Section V.A.2 on page 9 when the connectivity property of financial models as streams is explained;

- **Modifiers**: modifiers support the plasticity property of financial models, as extensively explained in Section V.A.2 on page 9. The plasticity function was formalized in Equation 4. A practical example of plasticity applied to finance was described in Figure 6;

- **Reactives**: reactives allow the representation of declarative, non-sequential properties in financial models: inverted control, abstraction of time management, and abstraction of synchronicity details. The formalization of reactive processors was given previously in Section V.B on page 10.
- **Agents**: this specialization of a processor is used to support a sub-classification of event-driven simulation model called agent emulation. Agent emulation is described in details in Section V.D.1 on page 14. On that section, an agent processor is described as software agent $A_i$, represented by stream $\phi(P, \delta)$, in Figure 14.

*C. Endpoints*

The third type of contributions in a computational representation for the field of economics is called an endpoint. Endpoints can be used as either a source or a destination of data fragments $x$ in the execution stream $\phi(P_i, \delta_i)$, as previously represented in Figure 5 on page 8.

Since endpoints are contributions, they follow what we call evidential properties of contributions, explained in Section II.B. As such, endpoints can be defined and reused across different financial models (execution streams, as explained in Section V.A.2), by different users, whenever that same endpoint, or state of data, is required. Depending on the intended use of the data, endpoints can be further classified as visualizations or datasets.

Visualizations can be static or dynamic. Static visualizations, called plots, show a complete and immutable representation of samples $(x_i ... x_j)$ in which the window associated to the interval $[i, j]$ is constant. On the other hand, dynamic visualizations - also referred to as animations - represent mutable windows, or samples, of data. Dynamic visualizations adjust a geometric representation in real time, depending on the arrival of new data.

The second type of endpoints is called a dataset. Datasets are a repository of transformed data fragments, as previously described in Figure 5, as either an entry point of virtually infinite streams of data fragments $x$, or a destination of the execution of stream $\phi(P_i, \delta_i)$.

Datasets can serve as intermediary entry and exit points of multiple sub-graphs or streams $\phi(P_i, \delta_i)$. In that sense, the resulting dataset of one execution stream can be a source dataset in a second, different, execution stream.

VII. CONSTRAINTS OF DATA

According to the representational process defined in Section II, the third element of a computational representation is called constraints of data. Constraints of data are explained in details in Section II.C, and define rules of association that establish what is feasible in a domain of knowledge. Those structural constraints use an abstract layer of data to define restrictions on a separate abstraction, based itself on data, hence the term meta-data. The set of structural constraints in a specific domain of knowledge is called meta-model.

For our domain of interest, financial sciences, structural constraints for associations between contributions and facets are defined in three of different groups of meta-models, based on its particular use: configuration, execution or simulation meta-model.

- **Configuration meta-model**: represents a versioned snapshot of a configuration of facets, arranged in a graph, over time. In other words, a structural description of all graphs defining the execution steps of a specific financial model. Since execution flows, or graphs, can change over time, a versioned configuration meta-model allows the exact definition and reproducibility of execution flow, at any given moment in the past. Instances of this meta-model will determine a reproducible sequence of execution, versions and provenance tracking of all data used to generate any specific result set.
- **Execution meta-model**: represent fragments of hierarchical data that flow through one or more compatible steps of a model. Instances of an execution meta-model are related to one specific configuration meta-model. In a sense is a description, in structured data, of concepts inherent to financial sciences: entities, contracts, instruments, or relationships [8].
- **Simulation meta-model**: supports the registration of experiments, results, and methods required to support an investigation. The registration is a permanent ternary association between financial models, shocks, and benchmarks. A financial model is a contribution describing the problem domain, the hypothesis under test, and the method under verification. The background for the definition of the hypothesis under tests and methods are part of the proof pipeline [5]. Shocks describe each of the executions of a financial model, used for recording utilized data, and results of each individual execution. Benchmarks describe the final comparison of results, of different shocks, and outline conclusions [8].

It is important to note that meta-models are defined and dependent on a finance case of use, or exercise, and should be defined in an ad hoc fashion, as required. To define an extensive set of meta-models that could be used in a large number of financial use cases is not practical, and would yield no additional insights to justify the increased complexity.

Additionally, some cases of use might require a partial set of meta-models. For example, for a real-time stock pricing financial model, given a strict dependency on mid-prices and a static price calculation function, a simulation and a configuration meta-model would not be necessary. A financial model for this specific case of use can rely exclusively on an execution meta-model [8].

VIII. CONCLUSION

Computers offer a number of overlapping and redundant ways to represent ideas, mostly because that is an unintended consequence of the need to support multiple possible representations across different domains of knowledge.

This research understands that part of an effective use of computational resources is to be able to properly formalize a domain of knowledge and allow to describe all concepts that would better fit that specific domain.

This is the intent of this paper in relation to economics: to formalize an effective computational representation for financial models in the field of economics.

We define a specific computational representation by defining a *knowledge representation system* [75] [19] [17] in terms of *what* can be shared, called in the scope of this research *contributions*, *how* to establish fundamental building blocks called *facets*, and *structural constraints* defined by constraints of data.

Facets define the computational representation in the framework. Combinations of those facets will serve as fundamental building blocks to other more complex abstractions in the conceptual framework. The term *contribution* applies to artifacts produced by participants (users) and transferred, or contributed, to a wider community of users through a shared scientific support system. Constraints of data define structural constraints for associations between contributions and facets, as well as data descriptions of high-level concepts on a financial model.

A computational representation is a non-cognitive enabler for *crowd-based scientific investigation*. As the term suggests, an enabler defines what is required, or should be in place, to enable a scientific investigation to occur across a large number of participants in a crowd [6] [76]. There are two types of enablers: cognitive and non-cognitive. Cognitive enablers are domain independent, or in other words, should be the same regardless of the characteristics of the domain of knowledge under consideration. This research considers two cognitive enablers: *methods of proof* and *collaboration in large-scale* [6]. The non-cognitive enabler, on the other hand, is strongly dependent on the specifics of a domain of knowledge. The representational process defined in Section II accounts for the strong dependency between domain-specific requirements and a computational representation.

A computational representation is, by definition, designed to evolve and adapt as new features of a domain of knowledge are introduced or brought into scope [6]. Therefore, a computational representation is never final. The computational representation described in this paper, Sigma, is not an exception. Sigma is intended to evolve as new financial models are investigated, and their requirements are brought in for study.

## IX. LIST OF EQUATIONS

## X. LIST OF FIGURES